\documentclass[prl,twocolumn,aps]{revtex4}
\usepackage{epsfig}

\begin{document}

\title{Shear-Transformation-Zone Theory of Glassy Diffusion, Stretched Exponentials, and the Stokes-Einstein Relation }

\author{J.S. Langer}
\affiliation {Department of Physics, University of California, Santa Barbara, CA  93106-9530}

\date{\today}

\begin{abstract}
The success of the shear-transformation-zone (STZ) theory in accounting for broadly peaked, frequency-dependent, glassy  viscoelastic response functions is based on the theory's first-principles prediction of a wide range of internal STZ transition rates. Here, I propose that the STZ's are the dynamic heterogeneities frequently invoked to explain Stokes-Einstein violations and stretched-exponential relaxation in glass-forming materials.  I find that, to be consistent with observations of Fickian diffusion near $T_g$, an STZ-based diffusion theory must include cascades of correlated events, but that the temperature dependence of the Stokes-Einstein ratio is determined by an STZ-induced enhancement of the viscosity. Stretched-exponential relaxation of density fluctuations emerges from the same distribution of STZ transition rates that predicts the viscoelastic behavior.
\end{abstract}
\maketitle

Among the deepest challenges in glass physics is understanding the temperature dependence of the Stokes-Einstein ratio in the neighborhood of the glass temperature $T_g$.\cite{EDIGER-06, EDIGER-09, BARTSCH-06}  An apparently related phenomenon is the stretched-exponential relaxation of density fluctuations and other correlations.  Both phenomena have been cited as evidence for dynamic heterogeneities in glassy systems, which supposedly provide rapid diffusion paths relevant to the first case, and a variety of localized, relaxation environments in the second.\cite{EDIGER-00}  Here, I propose that these heterogeneities are the same as the shear-transformation zones (STZ's) that Bouchbinder and I  \cite{BL-11} invoked to understand the broad range of time scales observed in frequency-dependent viscoelastic response functions.\cite{GAUTHIER-04} 

As described in earlier publications \cite{FL-98,JSL-08,FL-11}, STZ's are localized, shearable, thermally activated, structural fluctuations that appear and disappear on time scales of the order of the $\alpha$ relaxation time $\tau_{\alpha}$.  In the presence of an applied shear stress, they are the places where driven shear rearrangements occur.  They must also be associated with the self diffusion of a tagged molecule.  At most times, such a molecule remains in its cage, surrounded by its neighbors. It participates in a rearrangement involving these neighbors only when a thermally activated structural fluctuation -- a shear transformation -- occurs at its location.  When that happens, the tagged molecule jumps some distance, perhaps an intermolecular spacing $a$, in some direction, while one or more of its neighbors jumps in the other direction. A sequence of such jumps becomes a diffusive trajectory. 

For present purposes, consider only systems in thermodynamic equilibrium near their glass temperatures, and their linear responses to small deviatoric shear stresses $s$. The basic physics of the STZ model in this limit is expressed by the master equation for the number densities $n_{\pm}(\nu)$ of STZ's oriented parallel or antiparallel to $s$:
\begin{eqnarray}
\label{ndot}
\nonumber
\dot n_{\pm}(\nu) &=& R(\pm s)\,n_{\mp}(\nu) - R(\mp s)\,n_{\pm}(\nu)\cr\\  &+& \rho(\theta)\,[(1/2)\,e^{-\,e_Z/\theta} - n_{\pm}(\nu)].
\end{eqnarray}
Here, the $R(\pm s)$ are the rates of internal STZ transitions, $\theta = k_B\,T$ is the  temperature in energy units, and $e_Z$ is the STZ formation energy. $\rho(\theta)$ is the super-Arrhenius rate factor that rapidly decreases toward zero as  $\theta $ decreases through the glass temperature $\theta_g$. Times are measured in microscopic units, perhaps molecular vibration periods. The symbol $\nu$ denotes the internal transition rate, $\nu \equiv 2\,R(0)$; and writing Eq.(\ref{ndot}) with this explicit $\nu$ dependence implies that a range of different values of $\nu$ will be relevant to the phenomena of interest.  Denote the probability distribution for the $\nu$'s by $\tilde p(\nu)$. 

To start, compute the steady-state plastic strain rate $\dot\gamma^{pl}$ by setting $\dot n_{\pm}(\nu) = 0$ in Eq.(\ref{ndot}). The result can be written in the form
\begin{eqnarray}
\nonumber
\dot\gamma^{pl}&=& \epsilon_0\,\int_0^{\infty} d\nu\,\tilde p(\nu)\,\left[R(s)\,n_{-}(\nu) - R(- s)\,n_{+}(\nu)\right]\cr \\&=& \epsilon_0\,\int_0^{\infty} d\nu\,\tilde p(\nu)\,{1\over \tau(\nu)}\,{\cal T}(s),
\end{eqnarray}
where
\begin{equation}
\label{tau}
{1\over \tau(\nu)}= {1\over \tau_{\alpha}(\theta)}\,\left({\nu\over \nu + \rho}\right), ~~~ {1\over \tau_{\alpha}(\theta)}\equiv \rho(\theta)\,e^{-\,e_Z/\theta},
\end{equation}
and, to first order in $s$, 
\begin{equation}
\label{calT}
{\cal T}(s)\equiv {R(s) - R(-s)\over R(s) + R(-s)} \approx {v_0\,s\over \theta}.
\end{equation}

There are several aspects of these equations that need emphasis.  First, they assume that $\epsilon_0$, the volume of the plastic core of a shear transformation in units of $a^3$, is independent of $\nu$. Second, Eq.(\ref{tau}) includes a definition of $\tau_{\alpha}$ that is not tied directly to the temperature dependence of the viscosity $\eta$; that is, the $\theta$ dependence of $\eta$ will not be simply proportional to $\tau_{\alpha}(\theta)$. Third, ${\cal T}(s)$ is the stress-induced bias between forward and backward transitions.  The fact that this quantity is equal to $v_0\,s/\theta$ (where $v_0 \sim a^3$) is the classic Stokes-Einstein formula.  In \cite{BLII-III-09}, this result was shown more generally to follow from the second law of thermodynamics, with $\theta$ equal to the thermodynamically defined effective temperature $\chi$.  A simple application of the preceding analysis to sheared systems, where the transverse diffusion constant is $a^2/\tau_{\alpha}$, confirms that $\chi$ is the same as the effective temperature determined by a fluctuation-dissipation relation, at least within the STZ theory. (See \cite{CUGLIANDOLO-11} and references cited there.)  Throughout the following, thermodynamic equilibrium implies that $\chi = \theta$.

In \cite{BL-11}, Bouchbinder and I wrote the $\nu$-distribution in the form
\begin{equation}
\label{pnu}
\tilde p(\nu) = {\tilde A\over \nu\,[(\nu/\nu^*)^{\zeta} + (\nu^*/\nu)^{\zeta_1}]},
\end{equation}
where $\tilde A$ is a normalization constant.  The large-$\nu$ behavior, $\tilde p \sim \nu^{-1-\zeta}$, corresponds to a Boltzmann distribution over barrier heights, roughly consistent with that found experimentally by Argon and Kuo.\cite{ARGON-KUO-80} We used $\zeta_1 \sim 1$, and $\zeta \sim 0.4$ in accord with  viscoelastic data for various materials near their glass temperatures. The cutoff rate $\nu^*$ was determined by noting that the internal transition rate $\nu$ cannot be slower than the rate of spontaneous structural rearrangements. Therefore, $\nu^* \sim \tau_{\alpha}^{-1}$, which, for systems near their glass temperatures, is a factor of about $10^3$ smaller than $\rho$. These first-principles estimates of $\tilde p(\nu)$ and $\nu^*$ were accurately confirmed by the experimental data.  

Equation (\ref{tau}) implies that there are two different limiting populations of STZ's.  There are ``fast'' STZ's, with $\nu \gg \rho$, for which the time scale $\tau(\nu) \approx \tau_{\alpha}$.  These STZ's may make multiple, back and forth transitions between their two states during their lifetimes.  There is also a substantial population of ``slow'' STZ's with $\nu \ll \rho$ and $\tau(\nu) \approx (\rho/\nu)\,\tau_{\alpha} \gg \tau_{\alpha}$, which make at most one transition before disappearing.  Both populations are important for determining the inverse viscosity:
\begin{equation}
\label{eta}
{1\over \eta} = {\dot\gamma^{pl}\over s} = {\epsilon_0\,v_0\over \theta\,\tau_{\alpha}} \int_0^{\infty} d\nu\,\tilde p(\nu)\,\left({\nu\over \nu + \rho}\right).
\end{equation}

Turn now to the statistical problem of self diffusion.  Here, instead of averaging over probabilities of single events as above, we must consider sequences of multiple hopping events. The statistical weight assigned to any such sequence depends on how far the molecule moves and how long it takes to get there.  Technically, the molecule executes  a ``continuous-time random walk'' (CTRW). \cite{MONTROLL-SHLESINGER-84,BOUCHAUD-GEORGES-90} Note that the model of STZ-enabled diffusion steps contains no information about dynamics on sub-molecular space or time scales.  In effect, it is coarse grained on the scale of the molecular spacing, and it ``knows'' nothing about fast vibrational motions other than that they carry the thermal noise that activates barrier-crossing transitions. Therefore, this model cannot describe $\beta$ relaxation or any details of how a molecule moves within the cage formed by its neighbors, or of how it escapes from that cage. 

The basic ingredient of the CTRW analysis is the one-step probability that, after waiting a time $t$, an STZ fluctuates into existence at the position of the tagged molecule, and the molecule jumps  a distance $z$.  (There is no loss of generality in projecting the diffusion mechanism onto one dimension.)  For an STZ with an internal rate $\nu$, this normalized one-step probability distribution is 
\begin{equation}
\psi(t,z,\nu) = f(z,\nu)\,{1\over\tau(\nu)}\,e^{-\,t/\tau(\nu)}, 
\end{equation}
where $f(z,\nu)$ is the probability of a jump of length $z$. As indicated, $f(z,\nu)$ may depend on $\nu$. The crucial assumption here is that the time $\tau(\nu)$ appearing in this formula is the same as the quantity defined in Eq.(\ref{tau}).  In both cases, $\tau(\nu)^{-1}$ is the average rate at which STZ's appear, make one net internal transformation (perhaps after multiple back-and-forth transformations), and then disappear.  This elementary fluctuation mechanism should be common to both plastic deformation and self diffusion.

The probability of any random walk consisting of a sequence of such steps is a product of space-time convolutions of factors $\psi(t,z,\nu)$. If the STZ events are uncorrelated -- a major assumption -- then each factor can be averaged independently over $\nu$.  The walk ends after a final time interval $t$ within which the molecule does not make a further jump; thus the complete path probability contains one factor 
\begin{equation}
\phi(t,\nu)={1\over\tau(\nu)}\,\int_t^{\infty}dt' \,e^{-\,t'/\tau(\nu)}= e^{-\,t/\tau(\nu)},
\end{equation} 
again included in the convolution integrals and averaged over $\nu$.

These space and time convolutions are converted into products by computing Fourier and Laplace transforms, in terms of which the multi-step walk probabilities can be summed to all orders. The self-intermediate scattering function $\hat F(k,t)$, i.e. the $k$'th Fourier component of the diffusion profile, is given by an inverse Laplace transform:
\begin{equation}
\label{Fhat-w}
\hat F(k,t) = \int_{\delta-\,i\,\infty}^{\delta +\,i\,\infty} {dw\over 2\,\pi\,i}\, {e^{w\,t/\tau_{\alpha}}\,\tilde K(w)\over 1 - \tilde J(k,w)},~~~\delta > 0,
\end{equation}
where
\begin{equation}
\label{Kdef}
\tilde K(w)= \int_0^{\infty} d\nu\,{\tilde p(\nu)\over w + \lambda(\nu)},~~\lambda(\nu) = {\tau_{\alpha}\over \tau(\nu)} = {\nu\over \nu + \rho}      
\end{equation} 
is the Laplace transform of the averaged $\phi(t,\nu)$, and 
\begin{equation}
\label{Jdef}
\tilde J(k,w) = \int_0^{\infty} d\nu\,\tilde p(\nu)\,{\lambda(\nu)\,\hat f(k,\nu)\over w + \lambda(\nu)}
\end{equation}
is the analogous transformation of $\psi(t,z,\nu)$. 

\begin{figure}[here]
\centering \epsfig{width=.45\textwidth,file=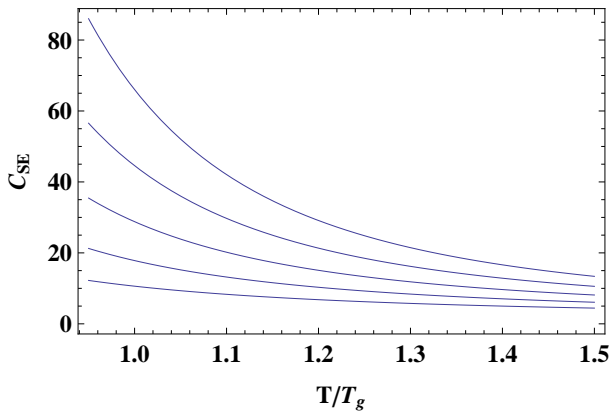} \caption{Stokes-Einstein ratio $C_{SE}$ for $\zeta$ = 0.4, 0.5, 0.6, 0.7, and 0.8 from bottom to top.} \label{CSE(T)}
\end{figure}

A crucial ingredient in these formulas is the jump-length distribution $\hat f(k,\nu)$. If, according to the discussion following Eq.(\ref{calT}), each STZ event moves the tagged molecule a mean-square distance $a^2$, then it would be natural to choose a Gaussian distribution $\hat f(k,\nu)= \exp(- k^2a^2/2)$ independent of $\nu$.  With this choice, however, the slow STZ's play the role of deep traps \cite{BOUCHAUD-GEORGES-90}, effectively shutting down long-time diffusion. To see this, compute the mean-square displacement $\overline{z^2(t)}$, which, for a $\nu$-independent $\hat f(k)$, becomes
\begin{equation}
\label{zsquared}
\overline{z^2(t)} = -\,\left[{\partial^2 \hat F(k,t)\over \partial k^2}\right]_{k=0}= a^2\,\int {dw\over 2\,\pi\,i}\, {e^{w\,t/\tau_{\alpha}}\over w^2\,\tilde K(w)}.
\end{equation} 
If $\tilde p(\nu)$ cuts off sharply at small $\nu$, then $K(w)$ is analytic at $w=0$, and the long-time behavior is obtained by integrating around the pole there.  The result is that $\overline{z^2(t)} \approx {\cal D}_0\,t$, with  ${\cal D}_0 = (a^2/\tau_{\alpha})\,\tilde K(0)^{-1}$.  Therefore, ordinary diffusion is suppressed by a factor $\tilde K(0)^{-1}\sim \nu^*/\rho \ll 1$. For $\zeta_1 \le 1$ in Eq.(\ref{pnu}), $\tilde K(0)$ diverges, and ${\cal D}_0$ vanishes.  This result is manifestly inconsistent with the experimental data reported in \cite{EDIGER-06, EDIGER-09, BARTSCH-06}, where Fickian diffusion was observed near the glass temperature.  

Given the $\tilde p(\nu)$ determined by viscoelasticity  \cite{BL-11}, this discrepancy between diffusion theory and experiment falsifies the postulate of uncorrelated short jumps. It seems, instead, to indicate that the cascades of correlated STZ ``flips'' seen in low-temperature numerical simulations (see \cite{CAROLI-LEMAITRE-11} and earlier papers cited there) are relevant to laboratory systems at temperatures near $T_g$.  A systematic generalization of the CTRW analysis to include strongly correlated sequences of jumps is beyond the scope of this investigation; but it is plausible that such sequences might appear approximately in the form of anomalously long jumps initiated by individual slow events. Unlike the fast STZ's, the slow ones are stable only in near-equilibrium glass formers at low temperatures, and they only occasionally undergo shear transformations during their lifetimes. Their small values of $\nu$ imply that their internal energy barriers are high; therefore, when they do make  transitions, they release substantial amounts of energy that could trigger subsequent events.  Suppose, as a first guess, that a slow event sends a tagged molecule on a random walk that lasts for a time $\tau(\nu)$, during which the jump rate is $\tau_{\alpha}^{-1}$. The mean-square displacement during this walk would be  $a^2\,\tau(\nu)/\tau_{\alpha}$, which is equal to $a^2$ for fast STZ's, but becomes substantially larger for slow ones.  

The corresponding jump-length distribution is
\begin{equation}
\label{fhat}
\hat f(k,\nu) = e^{- k^2a^2/2\,\lambda(\nu)} = e^{- k^2a^2 \,\tau(\nu)/2\,\tau_{\alpha}}.
\end{equation}
This choice of $\hat f(k,\nu)$ solves the problem of overly slow diffusion.  When it is used in Eq.(\ref{Fhat-w}), the quantity $\tilde K(w)$ in the denominator of Eq.(\ref{zsquared}) cancels out, and  $\overline{z^2(t)} = a^2\,t/\tau_{\alpha}$ at all times. This means that diffusion in this approximation is normal in the sense that ${\cal D} = a^2/\tau_{\alpha}$.  

With this expression for ${\cal D}$, and Eq.(\ref{eta}), the Stokes-Einstein ratio becomes 
\begin{equation}
\label{SE}
{a_0\,{\cal D}\,\eta\over \theta} = \left[\int_0^{\infty} d\nu\,\tilde p(\nu)\,\left({\nu\over \nu + \rho}\right)\right]^{-1} \equiv C_{SE}(\theta),
\end{equation}
where $a_0 \equiv \epsilon_0\,v_0/a^2$.  Since  $\tau_{\alpha}$ has cancelled out of Eq.(\ref{SE}), and $\tilde p(\nu)$ is a function of $\nu/\nu^*$, the only temperature dependence of $C_{SE}(\theta)$ occurs via the ratio $\nu^*/\rho \sim  \exp(-e_Z/\theta)$. In \cite{BL-11}, we estimated that $\exp(-e_Z/\theta_g) \sim 10^{-3}$.  That estimate has been used to plot the graphs of $C_{SE}$ as functions of $T/T_g$ shown in Fig. \ref{CSE(T)}.  The balance between the contributions of slow and fast STZ's in Eq.(\ref{SE}) is determined largely by the exponent $\zeta$ in Eq.(\ref{pnu}), thus $C_{SE}$ is shown in the figure for five different values of $\zeta$.  Note that this theory does not (yet) have enough microscopic physical content to make a  connection between low and high-temperature behaviors, or even to make material-specific predictions of the shape of $\tilde p(\nu)$ or the length scale $a_0$. We know that, at higher temperatures or in noisier driven situations, $\tilde p(\nu)$  becomes sharply peaked at much higher values of $\nu$, so that $C_{SE}\cong 1$.  Thus, the values of  $C_{SE}(\theta_g)$ in Fig. \ref{CSE(T)} imply Stokes-Einstein violations of roughly the magnitude seen experimentally, arising from the contributions of the slow STZ's to the value of the viscosity according to Eq.(\ref{eta}).  

\begin{figure}[here]
\centering \epsfig{width=.45\textwidth,file=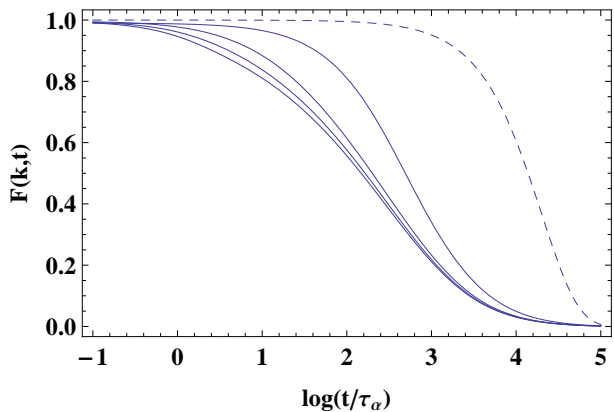} \caption{Computed self-intermediate scattering functions for $ka$ = 2.0, 1.0, 0.5, and 0.1, from left to right. The corresponding stretched-exponential indices are $b \cong$  0.43, 0.45, 0.58, and 0.78. The dashed curve is the Fickian limit for $ka = 0.01$. Only $\alpha$ relaxation appears here.  Space and time scales relevant to $\beta$ relaxation have been averaged out.} \label{SER-1}
\end{figure}

Finally, consider the self-intermediate scattering function $\hat F(k,t)$ given by  Eqs.(\ref{Fhat-w}-\ref{Jdef}).  To evaluate $\hat F(k,t)$ numerically, close the contour of integration in Eq.(\ref{Fhat-w}) around the branch cut on the negative real $w$ axis.  Some results are shown in Fig.\ref{SER-1}.  The parameters correspond roughly to a metallic glass near $T_g$: $\rho/\nu^* = 10^3$, $\zeta= 0.4$, and $\zeta_1 = 1$.  The wavenumbers are $k\,a = 2.0,\,1.0,\,0.5,\,{\rm and}\, 0.1$, from left to right.  All four curves exhibit stretched-exponential relaxation of the form $\exp\,[- \,c\,(t/\tau_{\alpha})^b]$, with $b \cong 0.43,\,0.45,\,0.58,\,{\rm and}\,0.78$, in the same order of decreasing $k$. 

Several features of these results are notable. First, there is a Fickian limit at small $k$, where $b$ approaches unity, and $\hat F(k,t) \sim \exp\,(-\,{\cal D}\,k^2\,t/2)$ as shown by the dashed line in Fig.\ref{SER-1} for $ka=0.01$.  Second, there is a non-Fickian limit at large $k$, where the curves begin to lie on top of each other, and $b$ approaches $0.4$.  This limiting behavior occurs because any choice of the factor $\hat f(k,\nu)$ vanishes at large $k$, leaving the initial, $k$-independent function $\tilde K(w)$ in the numerator of Eq.(\ref{Fhat-w}) as the only contribution to the relaxation function.  Note that 
\begin{equation}
\label{Kdef2}
K(t) \equiv \int {dw\over 2\,\pi\,i}\, e^{w\,t/\tau_{\alpha}}\,\tilde K(w)= \int_0^{\infty} d\nu\,\tilde p(\nu)\,e^{-\,t/\tau(\nu)}
\end{equation}
has the form of a local relaxation function averaged over a distribution of environments with different relaxation times $\tau(\nu)$. The transition between large-$k$ and small-$k$ behavior is temperature dependent {\it via} the quantity $\lambda(w)$ in Eq.(\ref{Fhat-w}), which produces a $\theta$ dependence here in much the way it did for the viscosity in Eq.(\ref{eta}).  At fixed $k$, according to these equations, the system changes from stretched-exponential to Fickian as the temperature increases.  It is not a coincidence that, in the large-$k$ limit given in Eq.(\ref{Kdef2}) and at low temperatures, $b \cong \zeta$; but $b = \zeta$ is not even an exact result for all choices of $\zeta$ and $\zeta_1$, nor is it exact in the limit of long times.

The analysis presented here calls for a unified investigation of Stokes-Einstein violations, stretched-exponential relaxation, and frequency dependent viscoelastic response functions, all in comparable glass-forming liquids near their glass temperatures.  There are major uncertainties, the most serious being the assumed $\hat f(k,\nu)$ in Eq.(\ref{fhat}).  Without some such assumption, the existence of the slow STZ's deduced from the viscoelastic experiments would be strongly inconsistent with the observations of Fickian diffusion near $T_g$. \cite{EDIGER-06, EDIGER-09, BARTSCH-06} However, it is not clear that Eq.(\ref{fhat}) is a realistic approximation, or whether a more accurate description of correlated diffusion steps might modify the theoretical Stokes-Einstein ratio.

More generally, checking the consistency of the various formulas shown here would test the theory as a whole and its interpretation of dynamic heterogeneities.  For example, it ought to be possible to measure viscoelastic responses (or internal-friction functions) using the same materials, under the same conditions, that are used for measuring $\hat F(k,t)$ in the large-$k$ limit shown in Eq.(\ref{Kdef2}).  Then, it should be possible either to back out consistent approximations for $\tilde p(\nu)$, or else to falsify the predicted consistency.  Similarly, it would be useful to make the latter kinds of measurements for the same systems in which Fickian diffusion is observed, and thus to test the approximation for $\hat f(k,\nu)$ in Eq.(\ref{fhat}).  Detailed studies of the $k$ and $\theta$ dependences of $\hat F(k,t)$, again on the same systems for which other measurements are being made, might further resolve the outstanding issues. 

\begin{acknowledgments}

I thank Eran Bouchbinder, Michael Cates, Mark Ediger, and Takeshi Egami for helpful discussions during the course of this investigation. This work was supported in part by the Division of Materials Science and Engineering, Office of Basic Energy Sciences, Department of Energy, DE-AC05-00OR-22725, through a subcontract from Oak Ridge National Laboratory. 

\end{acknowledgments}


\begin{thebibliography}{99}

\bibitem{EDIGER-06} M.K. Mapes, S.F. Swallen, and M.D. Ediger, J. Phys. Chem. B {\bf 110}, 507 (2006).

\bibitem{EDIGER-09} S.F. Swallen, K. Traynor, R.J. McMahon, and M.D. Ediger, J. Phys. Chem. B {\bf 113}, 4600 (2009). 

\bibitem{BARTSCH-06} A. Bartsch, K. Ratzke, F. Faupel, and A. Meyer, Appl. Phys. Lett. {\bf 89}, 121917 (2006). 

\bibitem{EDIGER-00} M.D. Ediger, Annu. Rev. Phys. Chem. {\bf 51}, 99 (2000). 

\bibitem{BL-11} E. Bouchbinder and J.S. Langer, Phys. Rev. Lett. {\bf 106}, 148301 (2011); Phys. Rev. E {\bf 83}, 061503 (2011).

\bibitem{GAUTHIER-04} C. Gauthier, J.-M. Pelletier, Q. Wand and J.J. Blandin, J. Non-Crystalline Solids, {\bf 345\&346}, 469 (2004).

\bibitem{FL-98} M. L. Falk and J. S. Langer, Phys. Rev. E, {\bf 57}, 7192 (1998).

\bibitem{JSL-08} J. S. Langer, Phys. Rev. E {\bf 77}, 021502 (2008).

\bibitem{FL-11} M. L. Falk and J. S. Langer, Annu. Rev. Condens. Matter Phys. {\bf 2}, 353 (2011).

\bibitem{BLII-III-09} E. Bouchbinder and J. S. Langer, Phys. Rev. E {\bf 80}, 031132 and 031133 (2009).

\bibitem{CUGLIANDOLO-11} L. Cugliandolo, arXiv:1104.4901 (2011).

\bibitem{ARGON-KUO-80} A. S. Argon and H. Y. Kuo, J. Non-Crystalline Solids {\bf 37}, 241  (1980).

\bibitem{MONTROLL-SHLESINGER-84} E.W. Montroll and M.F. Shlesinger, in {\it Studies in Statistical Mechanics XI}, edited by J.L.Lebowitz and E.W. Montroll (North-Holland Physics Publishing, Amsterdam, 1984), pp 1-121.

\bibitem{BOUCHAUD-GEORGES-90} J.-P. Bouchaud and A. Georges, Phys. Rep. {\bf 195}, 127 (1990).

\bibitem{CAROLI-LEMAITRE-11} J. Chattoraj, C. Caroli and A. Lemaitre, Phys. Rev. E {\bf 84}, 011501 (2011).

\end{thebibliography}
\end{document}